\documentclass{article}

\PassOptionsToPackage{square,numbers}{natbib}

\usepackage{PRIMEarxiv}

\RequirePackage{natbib}

\usepackage[utf8]{inputenc} 
\usepackage[T1]{fontenc}    
\usepackage{hyperref}       
\usepackage{url}            
\usepackage{booktabs}       
\usepackage{amsfonts}       
\usepackage{nicefrac}       
\usepackage{microtype}      
\usepackage{lipsum}
\usepackage{graphicx}
\graphicspath{{media/}}     

\usepackage{amsmath,amssymb,amsthm}
\usepackage{doi}
\usepackage{url}
\usepackage{nameref}
\usepackage{varioref}
\usepackage{hyperref}
\usepackage{cleveref}\usepackage[figuresright]{rotating}%
\usepackage{multirow}
\usepackage{longtable}
\usepackage{algorithm}%
\usepackage{algorithmicx}%
\usepackage{algpseudocode}%

\algrenewcommand\algorithmicrequire{\textbf{Input:}}
\algrenewcommand\algorithmicensure{\textbf{Output:}}
\def\sidewaystablefn{\renewcommand\footnotetext[2][]{{\removelastskip\vskip3pt%
\let\tablebodyfont\tablefootnotefont%
\hskip0pt\if!##1!\else{\smash{$^{##1}$}}\fi##2\par}}%
}%
\title{The Longitudinal Health, Income, and Employment Model (LHIEM): a discrete-time microsimulation model for policy analysis
\thanks{\textit{Journal of Artificial Societies and Social Simulation (2025).} }
}

\author{Adrienne M. Propp\\
ICME, Stanford University\\
RAND Corporation\\
\texttt{propp@stanford.edu} \\  
   \And
	Raffaele Vardavas \\
	RAND Corporation\\
    \And	 Carter C. Price \\
	 RAND Corporation \\
     \And
          Kandice A. Kapinos \\
	RAND Corporation \\
	UT Southwestern Medical Center \\
}

\begin{document}
\maketitle

\begin{abstract}
Dynamic microsimulation has long been recognized as a powerful tool for policy analysis, but in fact most major health policy simulations lack path dependency, a critical feature for evaluating policies that depend on accumulated outcomes such as retirement savings, wealth, or debt. We propose the Longitudinal Health, Income and Employment Model (LHIEM), a path-dependent discrete-time microsimulation that predicts annual health care expenditures, family income, and health status for the U.S. population over a multi-year period. LHIEM advances the population from year to year as a Markov chain with modules capturing the particular dynamics of each predictive attribute. LHIEM was designed to assess a health care financing proposal that would allow individuals to borrow from the U.S. government to cover health care costs, requiring careful tracking of medical expenditures and medical debt over time. However, LHIEM is flexible enough to be used for a range of modeling needs related to predicting health care spending and income over time. In this paper, we present the details of the model and all dynamic modules, and include a case study to demonstrate how LHIEM can be used to evaluate proposed policy changes.
\end{abstract}


\section{Introduction}\label{sec1}

The current policy landscape has brought about a number of proposed reforms to health care delivery in the United States (U.S.). To help guide decision-making, policymakers often turn to mathematical modeling to explore the potential outcomes of such reforms. Dynamic microsimulation is a powerful tool by which the predicted effects of a policy can be evaluated at multiple levels (e.g. individuals, households, and employers, as well as at the aggregate level) \citep{cite1, krijkamp2018microsimulation}.

We built the Longitudinal Health, Income, and Employment Model (LHIEM) to assess the impact of the 10Plan, an alternative health care financing proposal, on individuals, families and the federal government \citep{cite2}. The proposed policy involves a self-pay system designed to minimize the burden of health care costs by allowing individuals to borrow from the U.S. government to cover health care costs and repay them at a means-tested annual rate. The analysis thus required careful tracking of medical expenditures, medical debt, loan payments, and income over time. These requirements drove the design of LHIEM, a nationally representative dynamic microsimulation that evolves individual- and family-level health expenditures and key individual-level attributes, including labor force participation and income, health insurance status, family structure, and health status, over a multi-year period.
 
LHIEM builds on the strong foundation of dynamic microsimulation models in the health policy field, addressing a critical gap left by existing tools. In the U.S., five major health policy simulation models are widely used for federal and state-level decision making \citep{cite3}: the Health Insurance Simulation Model (HSIM, and more recently, HSIM2) developed by the Congressional Budget Office; the Gruber Microsimulation Model (GMSIM), developed by Dr. Jonathan Gruber at the Massachusetts Institute of Technology (MIT); the Comprehensive Assessment of Reform Efforts (COMPARE) model, developed by the RAND Corporation; the Health Benefits Simulation Model (HBSM), developed by the Lewin Group; and the Health Insurance Policy
Simulation Model (HIPSM), developed by the Urban Institute. While these partial equilibrium models are powerful tools, they are not designed to handle path-dependency, making them less suitable for evaluating scenarios where an individual’s healthcare payments depend on their history of expenses and payments. To analyze a policy such as the 10Plan, a model must explicitly capture the dynamic interplay between health spending, income, employment, family structure, and health status. Existing U.S.-based dynamic microsimulations, such as the Future American Model (FAM), are focused on different attributes (the progression of chronic health conditions, in the case of the FAM) rather than healthcare expenditures \citep{Fam1}.

Dynamic microsimulation models with some of the necessary features have been developed internationally --- for example, EUROMOD, which examines the impact of taxes on household incomes in Europe; Health Equity and its Economic Determinants (HEED), a European microsimulation for assessing the health impacts of income; and the Population Health Model (POHEM), which simulates the impact of government policies on health outcomes in Canada. However, the specific complexities of the U.S. healthcare system make these models challenging to adapt for analyzing U.S.-based policies. Nevertheless, these international models serve as important sources of inspiration for LHIEM.

LHIEM is thus distinctive as a dynamic model of healthcare spending and related attributes, offering a simpler and more customizable framework than many existing microsimulations. This simplicity makes it an accessible tool for other researchers to tailor to their specific research questions, with potential applications across health policy and related fields. The purpose of this paper is to describe the core technical components of LHIEM, rather than focusing on the specifics of the 10Plan policy proposal, to enable broader use of the model in exploring other research questions. LHIEM was developed in \textsf{R} and all code is available on \href{https://github.com/apropp/LHIEM}{GitHub}.

This paper begins with a description of the data and methods we used to develop LHIEM.
We continue with details on model validation and calibration, selected results from the analysis of the 10Plan \citep{cite2}, and uncertainty analysis. We conclude with a discussion of LHIEM's contributions, areas for future work, and how LHIEM fits in the context of microsimulation modeling for policy analysis.

\section{Methods}\label{sec2}
\subsection{Model overview}\label{subsec21}
\subsubsection{Purpose}\label{subsubsec211}

LHIEM facilitates the exploration of policy questions related to health care expenditures by estimating the evolution of medical spending and its key predictors over time for individuals and families. This discrete time microsimulation produces path-dependent life trajectories for each individual in our U.S.-representative model population by advancing the initial population from year to year as a Markov chain. Key individual-level attributes, including age, income, health status, health insurance status and source, and yearly medical spending are updated each year, based on the attributes of the previous year and transition rates derived from publicly available data. Individual records are also sampled each year for childbirth (if female), negative health shocks (such as an acute injury), the onset of a chronic disease, and death. These attributes and events contribute to determining an individual’s health care expenditures for the year.

\subsubsection{State variables and scales}\label{subsubsec212}
The model population consists of individuals organized into family units and is representative of the U.S. population on both the individual and family level. The set of state variables that describe individuals are included in Table~\ref{tab1}. Static state variables are set at model initialization or birth and do not change for the duration of an individual’s lifetime. Dynamic state variables may change at each time step of one year. Importantly, we assume that individuals do not interact with or affect one another. The only exception is with respect to income, which we aggregate to the level of the family unit.

\begin{sidewaystable*}
\sidewaystablefn%
\begin{center}
\begin{minipage}{\textheight}
\caption{Individual state variables}\label{tab1}
\begin{tabular*}{\textheight}{@{\extracolsep{\fill}}llllll@{\extracolsep{\fill}}}
\toprule%
\textbf{Variable group} & \textbf{Variable} & \textbf{Variable type} & \textbf{Units / Levels} & \textbf{Static v. Dynamic} & \textbf{Description} \\ \midrule
\multirow{5}{*}{Identification} & PID & String &  & Static & Simulation individual identifier \\
 & FID & String &  & Static & Simulation family identifier \\
 & TID & String &  & Dynamic & Simulation tax group identifier \\
 & WT & Numeric &  & Static & Individual weight \\
 & WTH & Numeric &  & Static & Household weight \\ \midrule
\multirow{4}{*}{\begin{tabular}[c]{@{}l@{}}Individual\\ characteristics\end{tabular}} & Sex & Factor & \begin{tabular}[c]{@{}l@{}}``Male",\\ ``Female"\end{tabular} & Static & Sex of individual \\
 & Age & Numeric & Years & Dynamic & Age of individual \\
 & Race & Factor & \begin{tabular}[c]{@{}l@{}}``Hispanic",\\ ``White",\\ ``Black",\\ ``Other"\end{tabular} & Static & Race of individual \\
 & Survive & Boolean &  & Dynamic & \begin{tabular}[c]{@{}l@{}}Whether individual is alive\\ in each year\end{tabular} \\ \midrule
\multirow{4}{*}{\begin{tabular}[c]{@{}l@{}}Economic\\ characteristics\end{tabular}} & FamIncome & Numeric & Nominal USD (\$) & Dynamic & Total family income \\
 & WageProp & Numeric & Proportion & Static & \begin{tabular}[c]{@{}l@{}}Proportion of family income\\ due to individual’s wages\end{tabular} \\
 & InsCat\footnotemark[1] & Factor & \begin{tabular}[c]{@{}l@{}}``Uninsured",\\ ``Medicaid",\\ ``Other Public",\\ ``NonGroup Private",\\ ``Other Private"\end{tabular} & Dynamic & Health insurance status \\
 & Deduct & Numeric & Nominal USD (\$) & Dynamic & Health insurance deductible \\ \midrule
\multirow{5}{*}{\begin{tabular}[c]{@{}l@{}}Health\\ characteristics\end{tabular}} & MedSpend & Numeric & Nominal USD (\$) & Dynamic & Medical spending this year \\
 & Visits & Numeric & Number of visits & Dynamic & \begin{tabular}[c]{@{}l@{}}Number of medical visits or\\ interactions this year\end{tabular} \\
 & Preg & Factor & \begin{tabular}[c]{@{}l@{}}``Pregnant",\\ ``Not Pregnant"\end{tabular} & Dynamic & Pregnancy status \\
 & HS & Factor & \begin{tabular}[c]{@{}l@{}}``Good",\\ ``Bad"\end{tabular} & Dynamic & Health status \\
 & Morbidity & Factor & \begin{tabular}[c]{@{}l@{}}``None",\\ ``Chronic",\\ ``Acute"\end{tabular} & Dynamic & Morbidity status \\ 
\bottomrule
\end{tabular*}
\end{minipage}
\end{center}
\footnotetext[1]{Note that Medicare is not included as a factor because this study was focused on the population under 65 years of age. Studies with different target populations would include this group.}
\end{sidewaystable*}

\subsubsection{Process overview and scheduling}\label{subsubsec213}
To properly characterize the evolution of medical spending, LHIEM evolves the dynamic attributes that we found to be the most significant predictors of medical spending. Specifically, these include:
\begin{itemize}
    \item Health status and morbidity,
    \item Family income,
    \item Mortality and end-of-life spending,
    \item Demographics, including pregnancy and population-level attributes.
\end{itemize}

LHIEM is implemented as a set of sequentially executed processes, which we call ``submodels.'' Each of the attributes listed above, as well as medical spending itself, is evolved by a separate submodel. These are described in detail beginning in paragraph \ref{submodels}. Algorithm~\ref{algo1} outlines the processes that comprise LHIEM in the order they are executed.

\begin{algorithm}
\caption{Model processes}\label{algo1}
\begin{algorithmic}[1]
\Require Model population in year $t$
\Ensure Model population in year $t+1$
\State update income
    \If{income changes by more than 10\%}\label{algln2}
        \State draw for update to insurance status
\EndIf
\State update FPL, accounting for inflation
\State update health status
\If{female}
    \State predict pregnancy
    \If{pregnant}
        \State estimate maternity costs
    \EndIf
\EndIf
\State update spending, calibrating to desired inflation rate
\State predict mortality
\If{deceased}
    \State inflate medical spending for last year of life
\Else
    \State update age
\EndIf
\State add newborn population
\State add net immigrant population
\end{algorithmic}
\end{algorithm}

\subsection{Model details}\label{subsec22}
\subsubsection{Initialization of model population}\label{subsubsec221}
An essential component of any microsimulation such as LHIEM is the model population. The LHIEM model population is constructed from census records and other individual-level sample data and is designed to mirror the real population's distribution over certain attributes that are relevant for the desired analysis.

We constructed LHIEM's initial model population primarily from the 2019 Annual Social and Economic (ASC) supplement to the Current Population Survey (CPS) \citep{CPS_ASEC_2019}. The CPS is a nationally representative survey containing most of the variables required for the model, including demographics (e.g. age, race, and sex), income (e.g. wages, salary, and other income sources), family structure (e.g. number of children and adults in each household), and health (e.g. self-reported health status and insurance source). We augmented this CPS-based model population with specific attributes from other datasets.

In particular, we used the 2015-2016 Medical Expenditure Panel Survey (MEPS) Panel 20 Longitudinal Data File \citep{MEPS2015_2016} to inform health care spending dynamics, one of the most important elements of LHIEM. The MEPS data contains information about individuals’ medical expenditures over the two-year period 2015-2016, as well as many of the other variables found in the CPS. We used the 2016 Medical Expenditure Panel Survey (MEPS) Person Round Plan (PRPL) file, drawn from the 2016 MEPS Household Component, to determine coverage type (single or family), plan metal level,\footnote{In the Health Insurance Marketplace, plan metal level is a ranking system that determines how an individual and their health care plan split health care costs.} and annual deductible level. We assigned each record the average deductible by coverage type and plan metal level, and for records without any plan or coverage-level data we assigned the overall average deductible \citep{cite5, cite6}. Given the well-documented differences between National Health Expenditures Accounts (NHEA) estimates and MEPS estimates of health care expenditures, we adjusted all expenditure amounts to match the NHEA using a factor of 1.27, following the work of \citet{cite7}.

We mapped the medical spending distribution and average deductible by age group and insurance status from the MEPS to the CPS-based initial population. We selected age group and insurance status as the matching variables\footnote{That is, the variables we used to identify "similar" individuals in the two datasets, allowing us to augment one dataset with attributes from the other.} because classification and regression tree (CART) analysis revealed these to be the strongest predictors of medical spending (see paragraph \ref{sec:medspend}). The age groups considered were: 18 and under, 19-34, 35-49, and 50-64\footnote{We excluded individuals over 65 in our study because these individuals are covered by Medicare and are thus not impacted by health care policy proposals such as the 10Plan.}. We harmonized insurance status categories between the MEPS and CPS, arriving at the five groups given in Table~\ref{tab1}. The harmonization rules for insurance status and other variables are included in Appendix \Cref{tab3}.

To ensure that the mapping from the MEPS to the CPS resulted in realistic levels of variation in medical expenditures, we compared summary statistics of the initial model population's expenditures with those reported in the literature \citep{ahrqSTATISTICALBRIEF}. As shown in Table~\ref{tab:summary_stats}, comparison across percentiles of the spending distribution demonstrates a high level of agreement between the MEPS target population and the initial model population.

\renewcommand{\arraystretch}{1.2}

\begin{longtable}{lccccccc}
\caption{Comparison of Medical Expenditures for Target and Model Populations}
\label{tab:summary_stats}\\
\toprule
\multicolumn{1}{l}{} & \multicolumn{7}{c}{\textbf{Average Individual Annual Expenditure by Pctile of Spending Distribution}} \\
\cmidrule(lr){2-8}
\textbf{} & \textbf{Overall} & \textbf{Bottom 50\%} & \textbf{Top 50\%} & \textbf{Top 30\%} & \textbf{Top 10\%} & \textbf{Top 5\%} & \textbf{Top 1\%} \\
\hline
\endfirsthead

\multicolumn{1}{l}{} & \multicolumn{7}{c}{\textbf{Pctile of Spending Distribution}} \\
\cmidrule(lr){2-8}
\textbf{} & \textbf{Overall} & \textbf{Bottom 50\%} & \textbf{Top 50\%} & \textbf{Top 30\%} & \textbf{Top 10\%} & \textbf{Top 5\%} & \textbf{Top 1\%} \\
\hline
\endhead

\textbf{Target Population (\$)} & 5,006 & 276 & 9,735 & 15,057 & 33,053 & 50,077 & 110,003 \\
\hline
\textbf{Model Population (\$)} & 4,748 & 271 & 9,224 & 14,379 & 33,267 & 52,524 & 131,250 \\
\bottomrule
\end{longtable}

We also validated the other attributes of our initial model population against external statistics to ensure general representativeness. We confirmed that each 5-year age group represented between 7-10\% of the under-65 population \citep{NCHS, census_age_sex}, and that our model population had a representative distribution of insurance status and sex across age groups (Tables~\ref{tab:summary_stats_ins} and \ref{tab:summary_stats_sex}, with target population data pulled from \cite{census_age_sex}).

\renewcommand{\arraystretch}{1.2}

\begin{longtable}{lcccc}
\caption{Comparison of Number of Uninsured Individuals by Age Group for Target and Model Populations}
\label{tab:summary_stats_ins}\\
\toprule
\textbf{Age Group} & \textbf{<19} & \textbf{19-34} & \textbf{35-49} & \textbf{50-64}  \\
\hline
\endfirsthead

\textbf{Age Group} & \textbf{<19} & \textbf{19-34} & \textbf{35-49} & \textbf{50-64} \\
\hline
\endhead

\textbf{Target Population (millions)} & 4.53 & 10.38 & 7.78 & 5.78 \\
\hline
\textbf{Model Population (millions)} & 4.12 & 9.71 & 7.38 & 5.26 \\
\bottomrule
\end{longtable}

\begin{longtable}{lcccc}
\caption{Comparison of Percent Female Individuals by Age Group for Target and Model Populations}
\label{tab:summary_stats_sex}\\
\toprule
\textbf{Age Group} & \textbf{<19} & \textbf{19-34} & \textbf{35-49} & \textbf{50-64}  \\
\hline
\endfirsthead

\textbf{Age Group} & \textbf{<19} & \textbf{19-34} & \textbf{35-49} & \textbf{50-64} \\
\hline
\endhead

\textbf{Target Population (\%)} & 49.0 & 49.9 & 50.8 & 51.6 \\
\hline
\textbf{Model Population (\%)} & 48.9 & 49.7 & 50.7 & 52.0 \\
\bottomrule
\end{longtable}

\subsubsection{Input data}\label{subsubsec222}
As discussed above, we used the CPS and MEPS as the primary datasets for the initial model population. However, the development of each submodel required additional data sources to ensure the accurate evolution of the LHIEM model population over time, specifically with respect to health status, income, mortality, and demographic shifts. 
These additional data sources include:
\begin{itemize}
    \item Panel Study on Income Dynamics (PSID) data from 2005 to 2017 to inform changes in the income distribution;
    \item U.S. Bureau of Labor Statistics CPI Inflation Calculator \citep{BLS_CPI_Calculator} to convert all dollar values to 2019 real dollars;
    \item Centers for Disease Control (CDC) data (including 2016 National Center for Health Statistics data, the 2017 National Vital Statistics Report, and 2019 American Cancer Society data) to inform the updates to health and health-related characteristics, including fertility, health status, and mortality;
    \item U.S. Census Bureau population-level projections from 2016-2017 to inform demographic changes --- specifically, birth, death, and migration by year.
\end{itemize}
The specific details of how each supplemental data source was used can be found in the discussion of the relevant submodel in the following sections.

\subsection{Submodels}
In this section, we describe each of the dynamic submodels in detail. Each submodel is designed to evolve particular attributes (specifically medical expenditures, health status, income, insurance status, mortality, and childbirth) from year to year. While some parameters and transition rates are assigned at the cohort level (e.g. by age, race, or sex), all transitions occur at the individual level.\footnote{In other words, while many individuals may be assigned the same transition rate for a given attribute, the occurrence of that transition is determined independently for each individual.}\label{submodels}

In our framework, each submodel functions independently, with the exception of medical expenditures. This particular submodel is directly dependent on the outcomes of all other submodels and, in turn, influences the mortality submodel. Although there are real-world interactions between attributes such as health and economic outcomes or health and fertility, these interactions are not typically the most significant predictors of the attributes' evolution over time. In designing LHIEM, we aimed to avoid unnecessary complexity, ensuring that any additional model components were justified by a significant improvement in the accuracy of aggregate outcomes. The resulting independence of the submodels provides the overall model with a high degree of modularity and flexibility.

\subsubsection{Medical expenditures}
We constructed a submodel to predict annual individual-level medical spending in each year based on medical spending in the previous year and demographic and medically relevant characteristics. The distribution of individual annual medical expenditures is known to be semi-continuous and highly skewed; many individuals incur no health care expenditures while some incur very high health care expenditures. A number of methods can be used to account for this \citep[see, for example,][]{cite8}. We employed a widely-used two-part model: part one predicts the likelihood of incurring nonzero medical expenditures; part two, applicable for cases with positive outcomes in part one, estimates the actual amount of these expenditures \citep{cite9,cite10,cite11}.

In part one, we estimated the likelihood of incurring nonzero expenditures using a logistic regression based on age group, sex, health status, insurance category, race, income, pregnancy status, nonzero spending in the previous year, and total amount of spending in the previous year.

In part two, we projected medical expenditures from year to year conditional on a prediction of nonzero expenditures in part one. We first used a classification and regression tree (CART) to identify partitions of the population exhibiting meaningfully distinct medical spending patterns. Here, a “partition” refers to a group of records sharing a certain combination of predictor variables. We then defined a unique spending regression for each partition identified by the CART using a generalized linear model with an identity link function and the natural log-transform of health care expenditures. This type of approach --- using machine learning to partition the dataset, then applying classical approximation schemes such as polynomial approximation or regression to the resulting partitions --- has been proposed as an effective method for solving high-dimensional regression problems in general contexts \citep{tiffany}.\label{sec:medspend}

CART analysis revealed that the regression coefficients meaningfully differed between age groups and levels of medical spending in the previous year. Insurance category, sex, income, race, and health status were also generally important in the regression. The final regressions ultimately predicted the next year’s health expenditures using spending in the previous year, age group, insurance category, sex, income, race, and health status as predictor variables.

We excluded pregnant women from the training sample because the costs associated with pregnancy are not easily determined from the MEPS Longitudinal File. We instead modeled the costs associated with maternity health care separately. We also excluded records with incomplete data from our analysis to help address the issue of missing data due to death. However, this method may be problematic for applications specifically concerned with old age or chronic disease \citep{cite12}.

To capture the wide variation in maternity and childbirth costs, for example in the event of Cesarean section delivery or complicated birth, we generated these costs stochastically using a truncated log-normal distribution fit to reproduce statistics from \citet{cite13} (median \$5,123, lower bound \$835, upper bound \$26,850) and to statistically reproduce the high level of variability in the costs across different insurance statuses and states. We modeled medical spending in the first year of life as a Poisson distribution, Pois($\lambda$), with rate parameter:
\begin{align}
    \lambda=\omega^{2.5}*1000.
\end{align}
Here, $\omega$ represents health status on a scale from 1 (good) to 5 (poor). We selected this form for rate parameter $\lambda$ in order to roughly reproduce statistics from the National Conference of State Legislatures memo on the costs of prenatal care \citep{cite14}. While we associated maternity and childbirth costs with the mother’s record, we associated medical costs in the first year of life with the infant record.

Medical expenditures thus evolve over time as the underlying predictor variables (health status, income and insurance status, age) evolve over time. The remainder of this section describes how LHIEM captures the dynamics of these predictor variables.

\subsubsection{Health status}\label{subsubsec232}
We assumed that individuals could be in a state of good health or poor health, with those in poor health suffering from either an acute condition (from which recovery is possible) or a chronic condition (from which recovery is not possible). To estimate the hazard rates for acute and chronic health conditions, we consulted the CDC and National Vital Statistics for cause of death data and obtained the proportions of deaths attributable to common health conditions \citep{cite15,cite16,cite17,cite18}. We assumed that these rates also reflect the proportions of incidence rates by age and gender (we elaborate on the implications of these assumptions in Section 4). We used these hazard rates to represent the net transition from good to poor health. We then implemented a recovery rate from acute conditions that decreases with age, using a half-life that varies from 1 year at birth to 30 years at age 50.

Importantly, we do not explicitly model any correlation between the health statuses of individuals in the same household. While this correlation may be significant for both acute and chronic effects, this simplification does not affect LHIEM's ability to capture the aggregate dynamics of medical expenditures for individuals, families, and the general population over time.

\subsubsection{Income and insurance status}\label{subsubsec233}
Because we were interested in longitudinal changes in an individual’s income and insurance status, it was important to ensure that the income trajectories were realistic and that the relationship between insurance status and income was accurately preserved over time. We constructed individual income trajectories using the income mobility estimates from the Panel Study for Income Dynamics (PSID) for 2006-2018. We inflated income to 2018 dollars using the Consumer Price Index Research Series Using Current Methods (CPI-U-RS).

We analyzed the annual change in real income across the income distribution to obtain probabilities for transitions across income and labor force categories by age. In particular, we updated incomes based on income quintile, age group (18-24, 25-34, 35-49, and 50-64), and sex, then sampled from the distribution of annual income changes for the relevant CPS records. Based on some of the more extreme annual changes in income found in the PSID, we assumed that a decrease in income of 90\% or more between years (which occurred in 1-3\% of cases for most age and income groups) indicated job loss. We then randomly assigned these individuals to an income in the bottom 15th percentile of the income distribution in the next year. To model transitions from unemployment to employment, we randomly assigned individuals in the bottom 15th percentile to an income sampled from the transitions of those who moved out of the bottom 15th percentile in the PSID. For each projection year, we applied the appropriate period-specific CPI for historical years and estimated CPI values for future years to ensure that all income numbers are expressed in nominal dollars for the respective years.

For efficiency, we aggregated individual income trajectories into family-level trajectories prior to running the model, but individual-level trajectories could also be used. We assumed that all individuals retired at age 65, and adjusted family income according to the proportion of family income that was previously attributable to the retiring family member’s wages or salary. Note that this assumption could produce an underestimation of family income in many cases, especially considering the trend towards later retirement and the tendency for unearned income to increase as a proportion of total income as individuals age.

Insurance status was initialized from the CPS. We assumed that it remained unchanged except in the event of an income shock of 10\% or more in either direction. We did not make any assumptions about changes in the insurance status distribution over time, and simply sampled from the original distribution of insurance status by federal poverty level (FPL) in the event of such an income shock. This is a limitation of our approach, as there may be many scenarios in which an individual or family changes insurance status without a significant change in income. However, this affects only a small share of the total population \citep{cite19} and thus does not greatly impact the results. 

\subsubsection{Mortality and end-of-life expenditures}\label{subsubsec234}
A substantial portion of lifetime medical expenditures occur in the last few months of life, particularly for those with chronic health conditions \citep{cite20}, so the treatment of mortality and end-of-life spending is an important aspect of the model. Because the MEPS does not cover all end-of-life spending, we obtained age- and gender-specific average probabilities of death from the 2017 Human Mortality Database U.S. life tables, assuming a smooth transition to 2018 using UN life table projections \citep{HMD, UNlife}. Pulling from the findings of \citet{cite21} on the relationship between predicted mortality and end-of-life spending, we were able to introduce individual-level variability in these probabilities and obtain an indirect dependence on health status.   We then used these predicted mortalities to determine which individuals would be removed from the population each year, to match the total mortality rate to the U.S. Census Bureau 2017 National Population Projections Tables \citep{cite22}. From there, we used the findings of \citet{cite21} on end-of-life health care spending to inflate medical expenditures for those predicted to die.

\subsubsection{Demographic changes}\label{subsubsec235}
It is important for a model such as LHIEM to capture changes to the overall population size and composition over time. In addition to simulating death (as described in the previous section), LHIEM also simulates birth and migration to account for such changes.

Birth, in particular, is critical to model because it contributes to both medical spending and demographic changes. We model birth as a Bernoulli random variable, where each female record has some probability of giving birth each year. We assigned these probabilities according to CDC fertility statistics based on age and race \citep{cite23}. New infant records introduced to the population were assigned the same demographic characteristics as the mother, with gender randomly assigned with equal probability and health status randomly assigned based on MEPS statistics.

To capture the effects of immigration, we added families to the model population according to the appropriate projections from the U.S. Census Bureau 2017 National Population Projections Tables \citep{cite22}. We randomly sampled the appropriate number of families from those labeled as immigrant families in the CPS dataset.

\section{Results}\label{sec3}
This section begins with an analysis of LHIEM’s performance, focusing on validation and calibration. We then present a case study to illustrate the application of LHIEM for policy analysis, along with example results derived from this case study. We conclude with results from our uncertainty analysis.

\subsection{Model validation}\label{modelperf}
To validate LHIEM's accuracy in capturing the dynamics of medical spending and its predictors, we conducted multiple validation exercises using historical data.

For individual-level validation of medical expenditures, we adopted an 85/15 data split for training and testing the CART and regression models, a standard practice in predictive modeling to balance model training and validation needs. This split provided a robust training dataset, with 11,448 observations in part one and 7,734 in part two, while ensuring a sufficient test set (15\%) to assess model performance and generalizability.

In part one, we determined that the best performing model for forecasting non-zero medical expenses in a given year incorporated the following factors as covariates: age group, sex, health status, insurance status, pregnancy status, race, family income, a binary indicator of non-zero health expenditures in the previous year, and total health expenditures in the previous year. In cross-validation exercises, this model outperformed alternative models using subsets of the list of covariates identified.

\begin{figure*}[h!]
\centering
\includegraphics[width=0.8\textwidth]{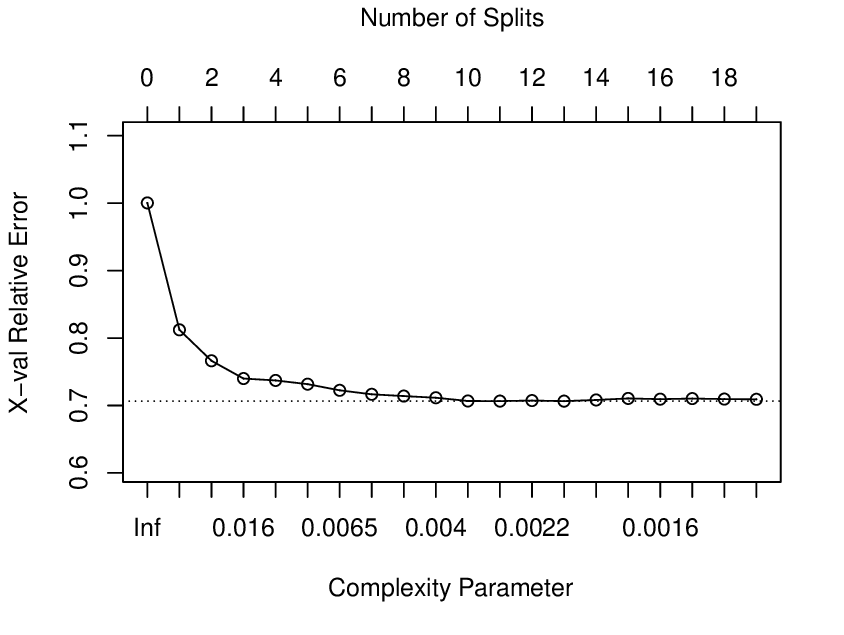}
\caption{Cross-validation relative error plotted against model complexity for regression trees analyzing nonzero medical expenditures. The y-axis represents the cross-validation relative error, and the top x-axis denotes the number of tree splits. The dashed line indicates the point of one standard deviation above the minimum error, guiding the pruning decision based on the "one standard error" rule.}\label{fig:err}
\end{figure*}

In part two, we performed the CART analysis using the following factors as covariates: age group, sex, health status, insurance status, race, family income, and total health expenditures in the previous year. We used the R library \texttt{rpart} to construct, optimize, and validate candidate CART models. This library provides tools for users to manage model complexity by setting a complexity parameter. This parameter defines the minimum required improvement in model fit that justifies further branching of the tree. A common strategy involves initially growing a regression tree beyond the desired complexity and then pruning it to retain only the most significant branches. The "one-standard error" rule, a widely adopted practice, guides this pruning process \citep{1se}. According to this rule, we trim the tree at the point where the cross-validation error is within one standard deviation above the minimum error observed.

The \texttt{plotcp()} function in R aids in this process by plotting the cross-validation error against the complexity parameter, highlighting the one-standard deviation threshold with a dashed line. The results of this validation process for the model of nonzero medical expenditures are depicted in Figure~\ref{fig:err}, which shows cross-validation error as a function of the number of splits (top $x$-axis) and corresponding complexity parameter (bottom $x$-axis).
According to the one standard error rule, the optimal complexity parameter is 0.0019, leading to a tree with 12 splits and 13 terminal nodes. However, the figure also indicates potential overfitting with larger trees, as gains in cross-validation error taper off quickly beyond the initial splits. This observation steered us to a conservative approach, opting to prune the tree down to six splits. This equates to a complexity parameter of 0.006, striking a balance between model complexity and predictive performance.

\begin{figure*}[h!]
\centering
\includegraphics[width=0.8\textwidth]{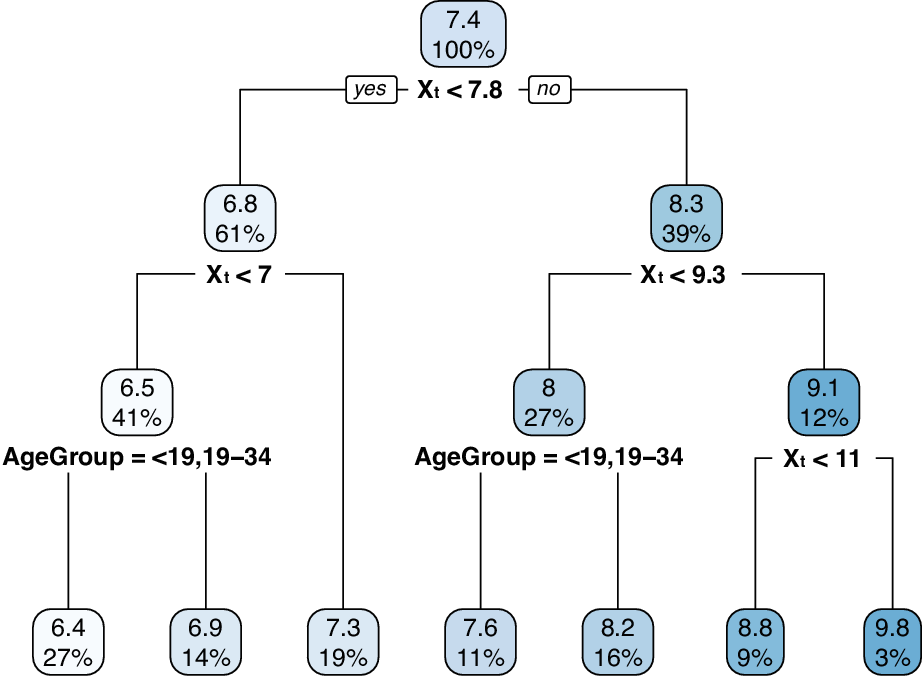}
\caption{Baseline CART output, where X$_t$ represents the log-transformed medical expenditures in year $t$. The tree illustrates the partitions remaining after pruning to achieve the desired complexity. Each box presents the predicted log-transformed medical expenditures for year $t+1$ (the outcome variable) at the top, and the percentage of the training data within each partition at the bottom. Labels below each box describe the condition for the split leading to the respective partitions. The final row of boxes represents the complete set of partitions used for analysis.}\label{fig:CART}
\end{figure*}

The CART resulting from this analysis is depicted in Figure~\ref{fig:CART}. In this diagram, X$_t$ indicates the natural log-transform of medical expenditures in year $t$. The top number in each box provides the CART estimate for $\text{X}_{t+1}$ (the outcome variable, the natural log-transform of medical expenditures in year $t+1$), and the bottom number provides the percentage of the training sample that falls into each partition. The six splits correspond to seven leaves (or partitions), each of a suitable sample size to estimate a regression. For each leaf in the resulting regression tree, we then built a separate regression using the same covariates and a generalized linear model assuming normally distributed errors.

Validation of aggregate model results against NHEA data revealed a higher-than-expected medical inflation rate of 6.9\% per year. We therefore added a calibration step to control the rate of inflation. We first normalized the year $t+1$ expenditures by dividing by the weighted inflation rate, then multiplied by the desired inflation rate. We pegged the inflation of medical spending to the projected growth in Medicare per capita spending, rather than to Medical CPI, and thus assumed a growth rate of 5.1\% \citep{cite26}. Medical expenditure estimates for each year of the simulation were then recalibrated to this inflation rate. This calibration step reflects the tendency for dynamic microsimulations to require alignment to aggregate benchmarks \citep{cite27}.

\begin{figure*}[h]
\centering
\includegraphics[width=\textwidth]{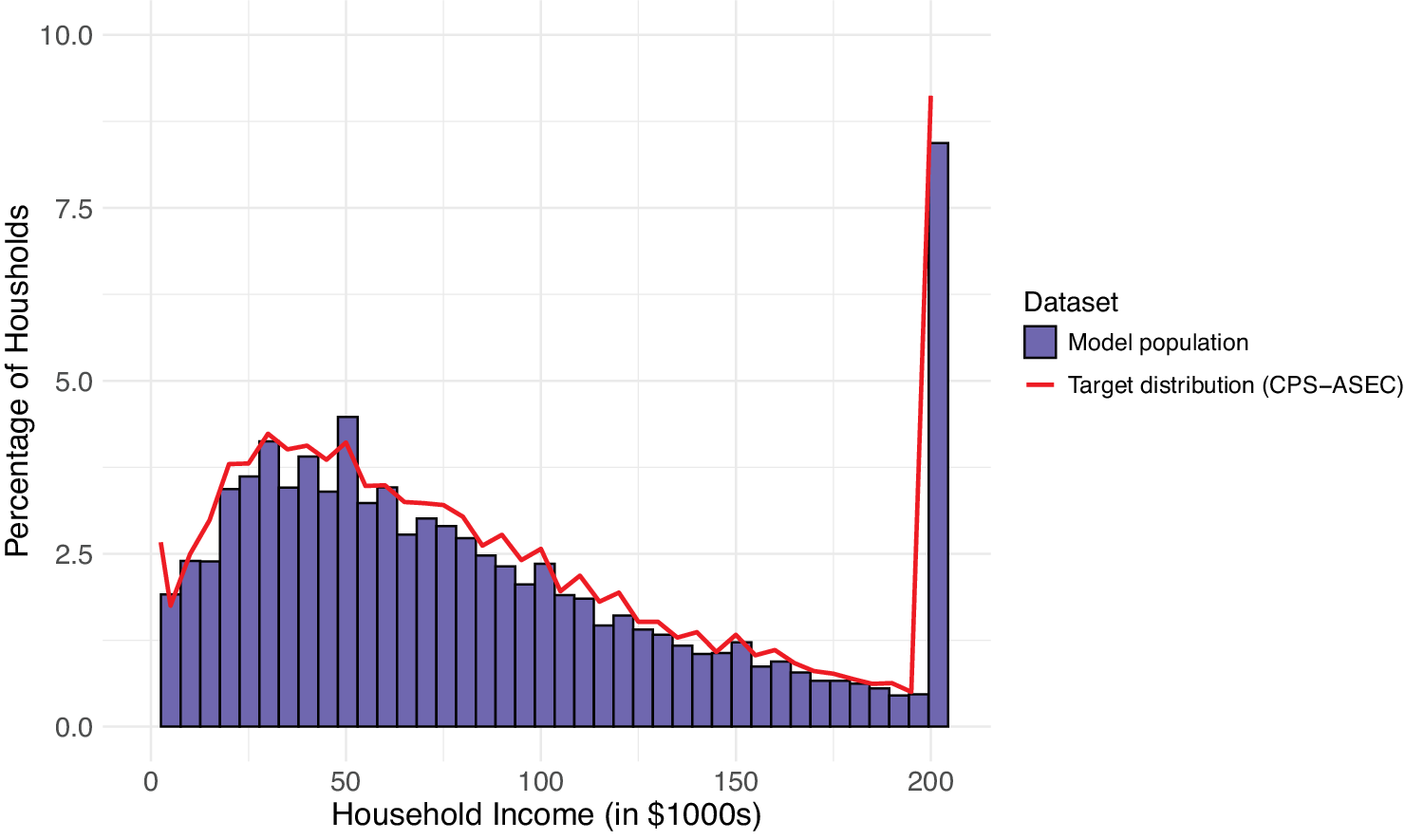}
\caption{Validation of income distribution against the Current Population Survey (CPS) Annual Social and Economic (ASEC) Supplement \citep{cps_income}. The last bin represents all households with incomes above \$200,000.}\label{fig:income_dist}
\end{figure*}

To validate the income submodel, we analyzed both the individual income trajectories and the overall income distribution generated by the model. We first ensured face validity, which involved subjectively assessing whether the model's income trajectories and distributions appeared realistic and plausible when compared to known economic patterns. Subsequently, for a more objective assessment, we verified that the distribution of changes in income from one year to the next aligned statistically with the 2006-2018 PSID data.Additionally, we compared the income distributions our model produced for years 1, 5, 10, and 15 against the corresponding distributions in the PSID dataset, checking for significant deviations. Our model consistently aligned with each validation criterion considered, affirming its robustness and reliability in realistically simulating income trajectories and distributions. \Cref{fig:income_dist} shows how the income distribution of the LHIEM model population compares to the income distribution according to data from the CPS-ASEC \citep{cps_income}.

\begin{figure*}[h]
\centering
\includegraphics[width=\textwidth]{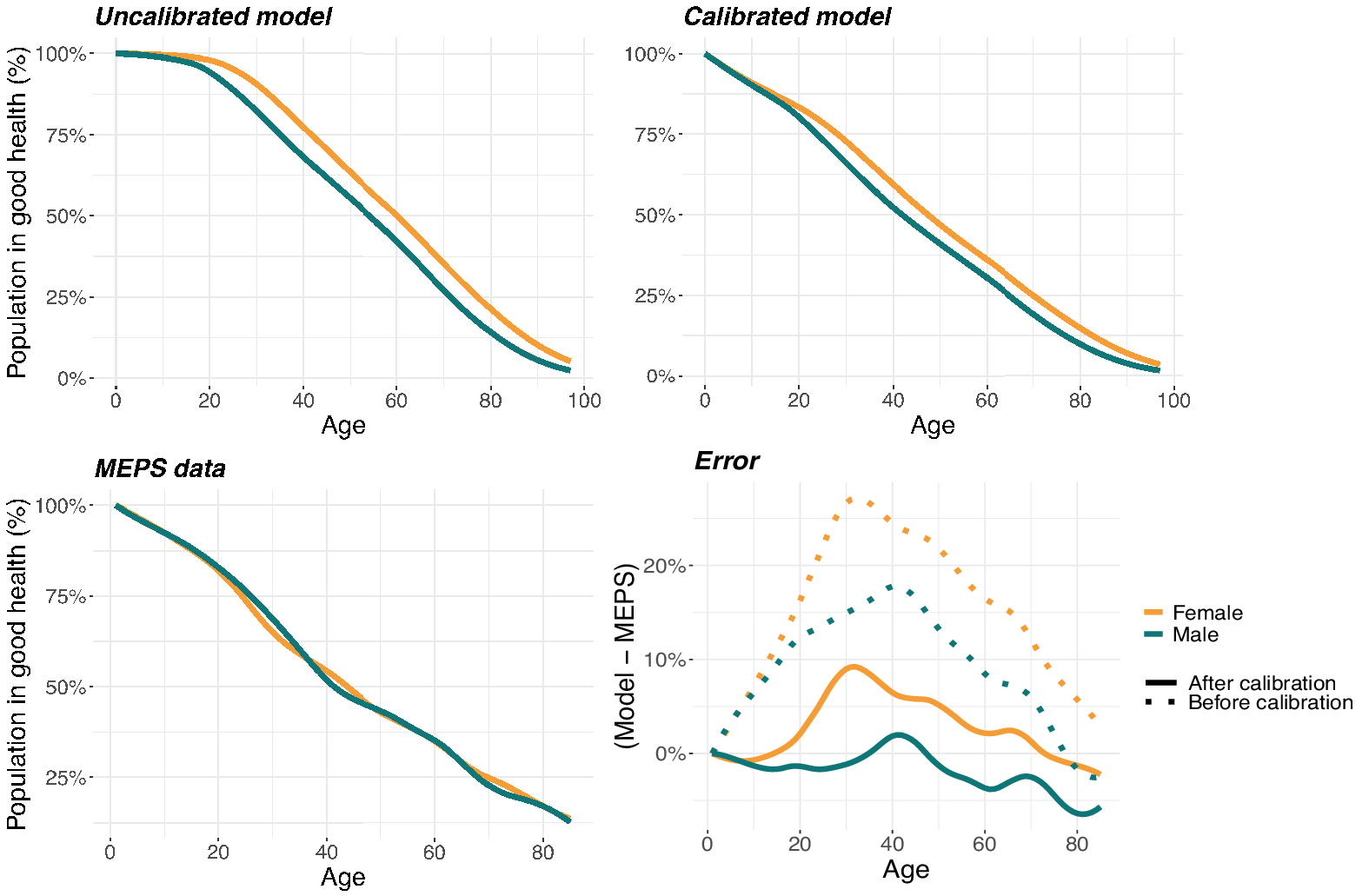}
\caption{Validation of health status submodel against MEPS data. Our uncalibrated health status module overestimated the incidence of ``good'' health status among younger individuals (upper left) relative to the MEPS population (lower left). Our model showed much better alignment with the MEPS after calibration (upper right), highlighting the importance of model validation and calibration. The lower right panel compares the error relative to the MEPS before and after calibration, where error is calculated as model proportion in good health minus MEPS proportion in good health.}\label{fig:healthstat}
\end{figure*}

Validation of the health status submodel was crucial, as transitions between good and poor health states --- whether driven by acute or chronic conditions --- are key predictors of medical expenditures. To ensure that the health status distributions by age and gender were accurately represented over time, we benchmarked our incidence rates for acute and chronic conditions against MEPS data. It is important to note that MEPS health status is self-reported and may not fully capture the true prevalence of conditions, unlike the CDC incidence
 data which informs the dynamics of our health status submodel. Indeed, initial comparisons revealed an overestimation of ``good'' health status among individuals under 25 years in our model. To address this discrepancy, we adjusted the incidence rates of acute conditions for this age group to better align with MEPS data. \Cref{fig:healthstat} shows the proportion of individuals in good health as a function of age for the uncalibrated model, the calibrated model, and the MEPS data used for comparison. Further analysis revealed that these adjustments significantly affected projections of individual and household healthcare expenditures, as shown in the results for Model 4 in \Cref{tab2}. This validation process underscores the importance of calibration in model validation to ensure both the accuracy and applicability of our findings.\label{sec:val_hs}

\subsection{Case study: an alternative health care financing approach}\label{sec4}
We developed LHIEM in order to analyze the 10Plan, a proposed alternative health care financing approach targeting individuals who are uninsured or purchase private coverage in the nongroup health insurance market \citep{cite2}. This policy introduces a self-pay system designed to reduce the financial burden of health care costs for eligible individuals and families.

For this analysis, LHIEM provided dynamic estimates of health care expenditures at the individual, family, and population levels, along with estimated federal costs of implementation. The model incorporated predicted changes in health care utilization and prices over a 15-year time horizon. To simulate the proposed policy, we developed an additional module for tracking loans and payments associated with the proposed financing system. Further information on this extension can be found in \cite{cite2}.

Importantly, we designed this policy module in a way that integrates seamlessly with LHIEM’s core model mechanics but easily allows for modification or replacement to explore other policy scenarios. This modular approach also allows us to distinguish between uncertainty arising from the core model mechanics and uncertainty related to the policy-specific elements.

We next present a sample of results from this analysis, focusing on the key outputs relevant to assessing the proposed policy’s impacts.

\subsubsection{Results}\label{subsec42}
Under the study’s baseline assumptions, our analysis suggested that the policy would cover approximately 46 million individuals (those who are currently uninsured or purchase private coverage in the nongroup health insurance market) and reduce total health care expenditures by \$33 billion over 15 years. On average, this translated to a decrease in out-of-pocket costs of \$1,343 per year for covered individuals, though out-of-pocket costs would actually increase on average for those who are currently uninsured due to an expected increase in utilization. Importantly, the results varied dramatically depending on the price levels that families and individuals covered by the policy would face, but were less sensitive to the model parameters and other policy details tested \citep{cite2}.

\begin{figure*}
\centering
\includegraphics[width=0.9\textwidth]{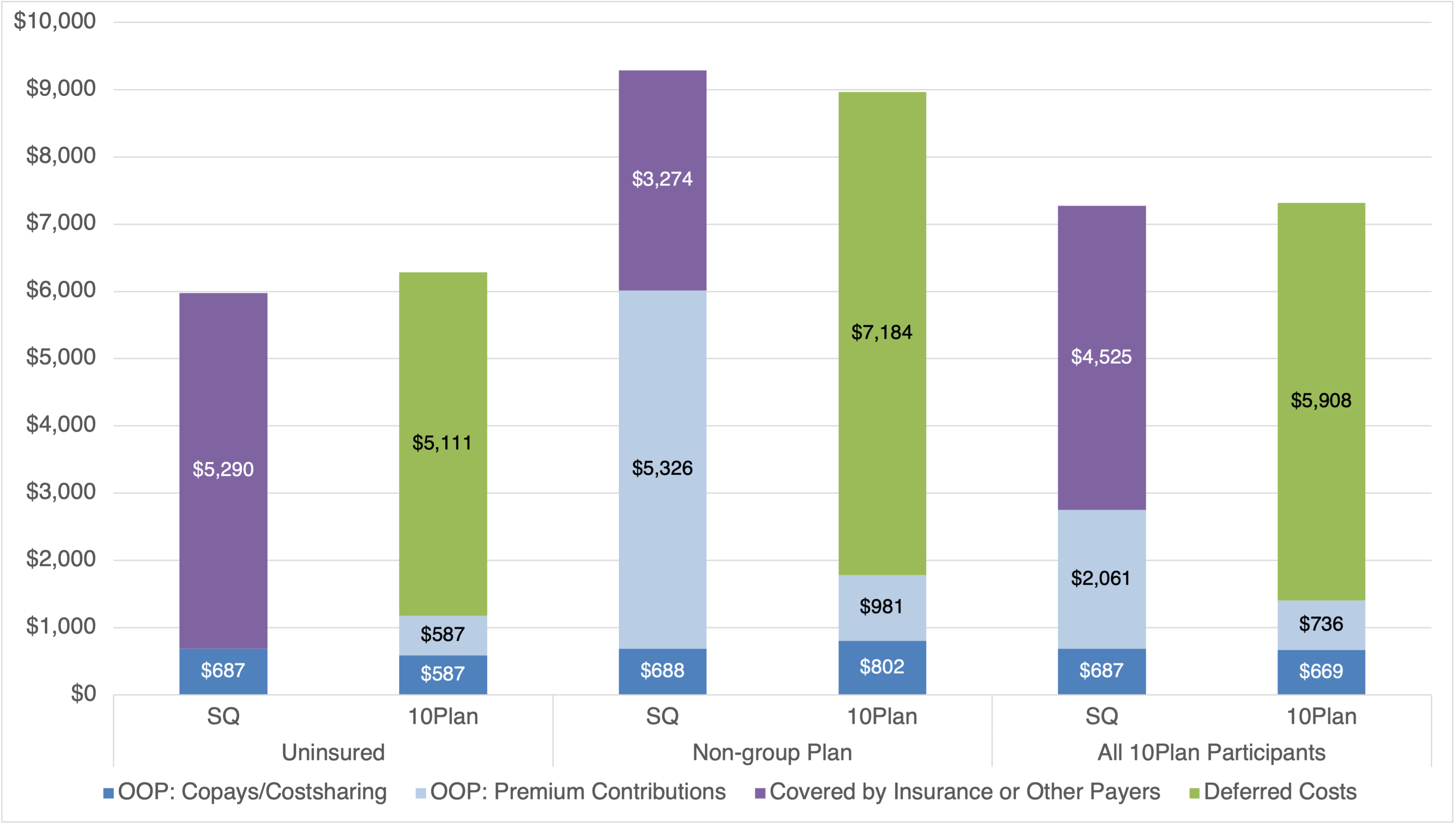}
\caption{Mean individual annual expenditures as of year 15 of the simulation, by current health insurance status. OOP costs are lower for most individuals under the proposed policy than under the status quo. The exception is for previously uninsured individuals, who would face higher OOP costs under the proposed policy}\label{fig:policy}
\end{figure*}

Figure~\ref{fig:policy} shows the estimated mean annual health care expenditures among individuals who would be eligible for the self-pay system, both under the status quo (SQ) and under the proposed self-pay system with baseline assumptions. The differences in total expenditures between the status quo and proposed policy scenarios can be attributed to assumptions about changes in utilization and health care prices under the self-pay system. Notably, individuals who are currently uninsured would face an increase in their out-of-pocket (OOP) costs, while individuals who are currently covered by a non-group plan would benefit from a substantial decrease in their out-of-pocket costs. On average, individuals eligible for the self-pay system would pay less out-of-pocket under the proposed policy than under the status quo.

\subsection{Uncertainty analysis}\label{secUncert}
Understanding the inherent uncertainties in a complex model like LHIEM is essential for ensuring its reliability and applicability. To address this, we conducted a comprehensive uncertainty analysis to examine the effects of both parameter uncertainty (stemming from incomplete information, imperfect data, and estimation error) and model uncertainty (arising from simplifications and assumptions in model formulation). 

\Cref{tab:models} outlines the core model variations tested in this analysis, each representing a modification to a specific parameter or assumption in the model mechanics. \Cref{tab2} summarizes the results for these variations, highlighting the percentage change from the baseline policy scenario. This comparison allows us to evaluate the impact of individual parameters or modeling assumptions on outcomes at the individual and household levels. Additionally, \Cref{tab:uncertainty_results_within} provides detailed summary statistics at the household level, offering a more nuanced view of the variation and uncertainty within each model.

Model 4, which explores the effect of the health status calibration step discussed in paragraph \ref{sec:val_hs}, yielded particularly revealing insights among the 7 model variants tested. This modification resulted in the greatest deviations from baseline results, highlighting a potential disparity between self-reported health status and actual medical conditions according to CDC statistics. Since LHIEM's medical expenditure predictions rely on self-reported health status, achieving alignment with the MEPS distribution was critical. This result underscores the importance of calibration in model validation and illustrates how uncertainty analysis can identify key factors that influence model performance and outcome accuracy.

In addition to the seven core model variations in \Cref{tab:models}, our full analysis included over 20 additional model variants focused on specific policy details and their effects on the healthcare landscape. These variants explored factors such as medical price levels, healthcare demand, loan repayment rates, and loan interest rates. While these policy-specific variations are not covered in this section, their results are discussed in detail in \cite{cite2}.

\renewcommand{\arraystretch}{1.4} 

\begin{table}[htbp]
\centering
\caption{Detailed Description of LHIEM Model Variants}\label{tab:models}
\begin{tabular}{p{3.7cm}p{2.0cm}p{9.9cm}}
\toprule
\textbf{Model variant} & \textbf{\mbox{Module affected}} & \textbf{Detailed description} \\
\hline
1. Status Quo & Policy & This model runs LHIEM with the policy module turned off. This represents the predictions produced by LHIEM without assuming any policy intervention.\\
2. Baseline policy & Policy & This model runs LHIEM with the policy module turned on. This represents the predictions produced by LHIEM under the baseline policy intervention examined. \\
3. Baseline policy with lower rate of medical inflation & Medical expenditures & The inflation rate for health care is separate from the overall inflation rate for the economy. This model tested the sensitivity of LHIEM long-term projections to the rate of medical inflation assumed. \\
4. Baseline policy without calibrating distribution of health status to MEPS & Health status & As described earlier, we calibrated LHIEM's distribution of health status to the MEPS self-reported health status distribution. By removing the calibration step, this model assesses the sensitivity of LHIEM to calibration of health status. \\
5. Baseline policy with end-of-life consumption inflated only for those with chronic conditions & Mortality and end-of-life expenditures & While significant medical expenditures are often incurred in the last year of life, the costs incurred by an individual suffering from a serious chronic condition (for example, cancer) will likely far exceed the costs incurred by the victim of a car accident, for example. This model assesses the sensitivity of LHIEM to inflating the end-of-life healthcare costs for only those individuals suffering from chronic conditions, rather than for all individuals. \\
6. Baseline policy with deflation of mortality scaling factor & Mortality and end-of-life expenditures & Our method for estimating the relationship between mortality and health care spending is based on the work of \cite{cite21}. To assess LHIEM's sensitivity to the characterization of this relationship, we tested a model that scales the estimated function relating mortality to health care spending. \\
7. Baseline policy with insurance category corrections & Income and insurance status & Since the CPS data on health insurance coverage is self-reported, LHIEM's initial assignment of health insurance status may contain some inconsistencies. To test the sensitivity of LHIEM to potential misreporting of insurance status, we performed the following steps: 1) reassign anyone on Medicaid and at or above 400\% FPL to have Group Private insurance; 2) reassign anyone on Medicaid and between 200-400\% FPL to have Nongroup Private insurance; 3) reassign individuals at or below 100\% FPL with Nongroup Private insurance to be Uninsured. \\
\bottomrule
\end{tabular}
\end{table}

\renewcommand{\arraystretch}{1.2} 

\begin{sidewaystable*}
\sidewaystablefn
\begin{center}
\begin{minipage}{\textheight}
\caption{Sample of simulation results: mean and median household and individual expenditures in year 5 and year 15 of the simulation, and corresponding percentage change from baseline for uncertainty analysis runs (Models 3-7)}\label{tab2}
\resizebox{\textwidth}{!}{
\begin{tabular}{llllllllllllllllll}
\toprule
\textbf{} & \multicolumn{5}{c}{\textbf{Mean expenditures (\$)}} &  & \multicolumn{5}{c}{\textbf{Median expenditures (\$)}} &  & \multicolumn{5}{c}{\textbf{\% Change from baseline median}} \\ 
\cline{2-6} \cline{8-12} \cline{14-18}
 & \multicolumn{2}{c}{\textbf{Individual}} &  & \multicolumn{2}{c}{\textbf{Household}} &  & \multicolumn{2}{c}{\textbf{Individual}} &  & \multicolumn{2}{c}{\textbf{Household}} &  & \multicolumn{2}{c}{\textbf{Individual}} &  & \multicolumn{2}{c}{\textbf{Household}} \\ 
\cline{2-3} \cline{5-6} \cline{8-9} \cline{11-12} \cline{14-15} \cline{17-18}
\textbf{Model Variant} & \textbf{Year 5} & \textbf{Year 15} &  & \textbf{Year 5} & \textbf{Year 15} &  & \textbf{Year 5} & \textbf{Year 15} &  & \textbf{Year 5} & \textbf{Year 15} &  & \textbf{Year 5} & \textbf{Year 15} &  & \textbf{Year 5} & \textbf{Year 15} \\ 
\midrule
1. Status quo & 4,414 & 7,274 &  & 5,554 & 8,960 &  & 2,279 & 3,421 &  & 3,162 & 4,689 &  &  &  &  &  &  \\
2. Baseline policy & 4,328 & 7,313 &  & 5,459 & 9,029 &  & 2,399 & 3,246 &  & 3,301 & 4,621 &  &  &  &  &  &  \\
3. \begin{tabular}[t]{@{}l@{}}Baseline policy with\\ lower rate of\\ medical inflation\end{tabular} & 4,301 & 7,306 &  & 5,418 & 8,961 &  & 2,388 & 3,261 &  & 3,178 & 4,651 &  & -0.45\% & 0.46\% &  & -3.72\% & 0.65\% \\
4. \begin{tabular}[t]{@{}l@{}}Baseline policy without\\ calibrating distribution\\ of health status to MEPS\end{tabular} & 4,325 & 7,276 &  & 5,445 & 8,965 &  & 2,328 & 2,995 &  & 3,034 & 4,296 &  & -2.96\% & -7.73\% &  & -8.09\% & -7.04\% \\
5. \begin{tabular}[t]{@{}l@{}}Baseline policy with end-of-life\\ consumption inflated only for\\ those with chronic conditions\end{tabular} & 4,343 & 7,266 &  & 5,500 & 8,945 &  & 2,398 & 3,219 &  & 3,305 & 4,584 &  & -0.06\% & -0.83\% &  & 0.13\% & -0.80\% \\
6. \begin{tabular}[t]{@{}l@{}}Baseline policy with\\ deflation of mortality\\ scaling factor\end{tabular} & 4,336 & 7,352 &  & 5,461 & 9,029 &  & 2,393 & 3,251 &  & 3,260 & 4,698 &  & -0.27\% & 0.14\% &  & -1.24\% & 1.67\% \\
7. \begin{tabular}[t]{@{}l@{}}Baseline policy with\\ insurance category\\ corrections\end{tabular} & 4,264 & 7,333 &  & 5,364 & 9,059 &  & 2,384 & 3,215 &  & 3,114 & 4,703 &  & -0.65\% & -0.97\% &  & -5.65\% & 1.79\% \\ 
\bottomrule
\end{tabular}
}
\end{minipage}
\end{center}
\end{sidewaystable*}

\renewcommand{\arraystretch}{1.2}

\begin{sidewaystable*}
\begin{center}
\begin{minipage}{\textheight}
\caption{Sample of simulation results: summary statistics for household-level medical expenditures in year 5 and year 15 of the simulation}
\label{tab:uncertainty_results_within}
\resizebox{\textwidth}{!}{
\begin{tabular}{lcccccccc}
\toprule
\multicolumn{1}{l}{} & \multicolumn{4}{c}{\textbf{Year 5 (\$)}} & \multicolumn{4}{c}{\textbf{Year 15 (\$)}} \\ 
\cmidrule(lr){2-5} \cmidrule(lr){6-9}
\multicolumn{1}{l}{\textbf{Model Variant}} & \textbf{Mean} & \textbf{25th Pct} & \textbf{50th Pct} & \textbf{75th Pct} & 
\textbf{Mean} & \textbf{25th Pct} & \textbf{50th Pct} & \textbf{75th Pct} \\
\midrule
1. Status quo & 5,554 & 1,298 & 3,162 & 7,254 & 8,960 & 1,649 & 4,689 & 11,146 \\
2. Baseline policy & 5,459 & 1,685 & 3,301 & 6,411 & 9,029 & 2,050 & 4,621 & 13,313 \\
3. \begin{tabular}[t]{@{}l@{}}Baseline policy with\\ lower rate of\\ medical inflation\end{tabular} & 5,418 & 1,707 & 3,178 & 6,378 & 8,961 & 2,026 & 4,651 & 12,941 \\
4. \begin{tabular}[t]{@{}l@{}}Baseline policy without\\ calibrating distribution\\ of health status to MEPS\end{tabular} & 5,445 & 1,657 & 3,034 & 6,316 & 8,965 & 1,979 & 4,296 & 13,274 \\
5. \begin{tabular}[t]{@{}l@{}}Baseline policy with end-of-life\\ consumption inflated only for\\ those with chronic conditions\end{tabular} & 5,500 & 1,721 & 3,305 & 6,397 & 8,945 & 1,985 & 4,584 & 13,003 \\
6. \begin{tabular}[t]{@{}l@{}}Baseline policy with\\ deflation of mortality\\ scaling factor\end{tabular}  & 5,461 & 1,720 & 3,260 & 6,408 & 9,029 & 2,053 & 4,698 & 13,216 \\
7. \begin{tabular}[t]{@{}l@{}}Baseline policy with\\ insurance category\\ corrections\end{tabular} & 5,364 & 1,688 & 3,114 & 6,300 & 9,059 & 2,035 & 4,703 & 13,407 \\
\bottomrule
\end{tabular}%
}
\end{minipage}
\end{center}
\end{sidewaystable*}

\section{Discussion}\label{sec5}
Investigating policies related to accumulated outcomes --- such as retirement savings, wealth, or bio-accumulative health processes --- requires models that account for path-dependent dynamics. In this paper we have introduced LHIEM, a dynamic microsimulation model capable of simulating medical expenditures, health status, income, mortality, and demographic attributes over time. By capturing these interrelated dynamics, LHIEM enables the analysis of path-dependent policies that cannot be adequately assessed using models that assume partial equilibrium. This expands the scope of policies that can be studied, providing valuable insights for policymaking across a variety of domains beyond health. 

LHIEM has already been applied to evaluate the 10Plan, an alternative healthcare financing approach \citep{cite2}. Specifically, the model was used to examine how total health care spending, individual- and family-level out-of-pocket health spending, and federal spending would change under a self-pay system aimed at reducing the financial burden for families without health insurance or those purchasing private coverage in the nongroup market. 

Our work builds on prior studies, publicly available datasets, and empirical evidence to create a modular and flexible dynamic microsimulation framework. A key strength of LHIEM is its reliance on open-source datasets and statistics that are regularly maintained and updated: the Current Population Survey (CPS); the Medical Expenditure Panel Survey (MEPS); the Panel Study on Income Dynamics (PSID); Centers for Disease Control (CDC) data; and U.S. Census Bureau population-level projections. This open data foundation ensures the reproducibility of our work and facilitates the use of LHIEM for future policy analysis.

Additionally, each of LHIEM's submodels is built using the most relevant data sources to accurately model the evolution of its respective dynamic variables. This modular design also supports targeted enhancements, enabling researchers to refine specific components to achieve greater accuracy or precision for particular variables of interest. Together, these features make LHIEM a robust and adaptable tool for studying complex path-dependent policy interventions.

\subsection{Areas for future work}
As with any simulation, we made several assumptions and simplifications, in some cases to improve model tractability and, in other cases, to avoid incorporating unnecessary sources of uncertainty and keep the model intuitive and straightforward. We have noted these throughout the paper. However, depending on the intended application, it may be desirable to modify particular aspects of the model to improve its fidelity for particular outcomes of interest. Here, we offer a discussion of potential areas for future work, and we hope that other researchers will expand or improve upon these where useful. \label{subsec51}

First, we note that LHIEM is a multi-year dynamic simulation constructed based on only two years of medical expenditure data. While it would have been preferable to develop a model of spending patterns with more than two years of data, the MEPS is not designed to link data across more than two years, and the authors are unaware of a publicly available dataset of comparable quality containing the required information for more than two years. The literature suggests that the relationship between medical spending in year $t$ and year $t+2$ is much weaker than that between spending in year $t$ and year $t+1$, so additional data of this nature would be unlikely to dramatically improve results \citep[see, for example,][]{cite29}. However, it may be worth exploring whether additional longitudinal data can improve predictive accuracy, particularly if richer datasets become available.

Another rich area for future work would involve refining our characterization of health status and its interaction with LHIEM's other submodels. First and foremost, the health transition rates would benefit from more detailed data directly capturing the hazard rates of various acute and chronic conditions, beyond the cause of death data we currently use. The calibration step aligning the LHIEM health status distributions to MEPS data currently corrects biases introduced by our approach, but higher-fidelity data could eliminate the need for this adjustment. In addition, a more direct dependence of end-of-life spending on health status could be appropriate. Currently, LHIEM admits an indirect dependence due to the assumed dependence of mortality rate on health spending and the dependence of health spending on health status. However, for a given mortality rate, it is possible that those suffering from chronic diseases may incur higher end-of-life costs than those with acute conditions. Furthermore, our scaling of the link between mortality and health spending found by \citet{cite21} and our application of this method to all age groups could introduce bias, as the relationship they found was actually based on the Medicare population. 

Additional extensions could address our assumption that pregnancy rates are independent of health status and health insurance status. In reality, women in poor health may not become pregnant at the same rate as their counterparts in good health. It is also well known that Medicaid pays for about 50\% of births, though this varies by state. Thus, we might expect some relationship between the likelihood of pregnancy and both health status and insurance status. Studies focusing specifically on maternal health, for example, may warrant a more detailed characterization of these relationships. These extensions would benefit from more detailed data on the relationships between medical spending, health status, pregnancy, and birth rates.

Finally, future studies should consider updating the data informing LHIEM's submodel dynamics. In particular, the income submodel is based on PSID data covering portions of the 2001-2007 business cycle, the 2008 recession, and the subsequent business cycle up to 2016. This period saw slower income growth and lower labor force participation in the U.S. relative to other periods. Thus, LHIEM projections would benefit from recalibration if macroeconomic conditions are expected to be better or worse than they were during this timeframe. For example, in the early months of the COVID-19 pandemic, we adapted this module to incorporate updated projections of unemployment, ESI responsiveness, and wage growth obtained from the Urban Institute \citep{cite30}. We recommend similar adaptations for future policy analyses, as well.

\subsection{Conclusions}\label{subsec52}
LHIEM is a novel dynamic longitudinal microsimulation developed to evaluate a proposed health care policy initiative. An important contribution of this work is the demonstration of the value of a flexible, modular approach; each submodel provides the necessary level of fidelity for a key component of the model and can easily be adapted based on the needs of the research question. In describing the model and data architecture in such detail, we aim to encourage further work in this area and facilitate this class of model development for future policy analysis.

LHIEM addresses a critical gap in health policy modeling by explicitly capturing path-dependency in healthcare expenditures. Unlike many existing models that treat each year independently, LHIEM explicitly tracks individual histories, making it particularly well-suited for evaluating policies where cumulative effects matter, such as healthcare financing reforms. LHIEM evolves key attributes like health status, income, employment, and medical expenditures over time, allowing for realistic life-course simulations. Its modular structure ensures flexibility, enabling researchers to adapt it to a wide range of policy questions, update components as better data becomes available, and extend its functionality as needed. LHIEM has already been validated against external datasets and applied to analyze the 10Plan, demonstrating its practical value in real-world policy evaluation.

While LHIEM represents a significant advancement, several areas warrant future development. Expanding the medical expenditure model beyond two years of data could enhance long-term projections, and updating transition probabilities would ensure alignment with changing economic and demographic trends. Improving end-of-life spending estimates and incorporating richer interactions between submodels, for example, between health status and fertility decisions, would further refine the accuracy of the model.

Given its flexibility and broad applicability, LHIEM is well-positioned for future research on healthcare financing, medical debt, and  other policy interventions that require detailed, longitudinal tracking of individual and household outcomes. As policy continues to evolve, a robust, adaptable modeling framework like LHIEM is essential for assessing long-term impacts and informing evidence-based decision-making. By enabling more nuanced, history-aware analyses, LHIEM provides policymakers and researchers with a powerful tool to explore complex hypothetical scenarios and craft sustainable, data-driven solutions to pressing policy challenges.
\section{Appendix A: Variable Harmonization}\label{secA1}

We used multiple datasets in the construction of LHIEM. In some cases, different datasets had different codifications of variables of interest. In these cases, it was necessary to harmonize across datasets. We include the harmonization rules we used for LHIEM in Table~\ref{tab3} below.

\renewcommand{\arraystretch}{1.2} 

\begin{longtable}{lllll}
\caption{Variable harmonization across datasets}
\label{tab3}\\
\toprule

Model variable & Data source & Data source variable name & Raw value & Harmonized value \\
\hline
\endfirsthead
\multicolumn{5}{c}%
{{\bfseries Table \thetable\ continued from previous page}} \\
\toprule
Model variable & Data source & Data source variable name & Raw value & Harmonized value \\ \midrule
\endhead
\cline{1-5}
\multirow{22}{*}{Race} & \multirow{17}{*}{CPS} & \multirow{13}{*}{RACE} & 100 & \multirow{4}{*}{White} \\
 &  &  & 802 &  \\
 &  &  & 803 &  \\
 &  &  & 804 &  \\ \cline{4-5} 
 &  &  & 200 & \multirow{9}{*}{Black} \\
 &  &  & 801 &  \\
 &  &  & 805 &  \\
 &  &  & 806 &  \\
 &  &  & 807 &  \\
 &  &  & 810 &  \\
 &  &  & 811 &  \\
 &  &  & 814 &  \\
 &  &  & 816 &  \\ \cline{3-5} 
 &  & \multirow{4}{*}{HISPAN} & \begin{tabular}[c]{@{}l@{}}All values except\\ the following:\end{tabular} & Hispanic \\
 &  &  & 0 &  \\
 &  &  & 901 &  \\
 &  &  & 902 &  \\ \cline{2-5} 
 & \multirow{5}{*}{MEPS} & \multirow{5}{*}{RACETHX} & 1 & Hispanic \\
 &  &  & 2 & White \\
 &  &  & 3 & Black \\
 &  &  & 4 & \multirow{2}{*}{Other} \\
 &  &  & 5 &  \\ \cline{1-5}
\multirow{26}{*}{InsCat} & \multirow{13}{*}{CPS} & ANYCOVLY & 1 & Uninsured \\ \cline{3-5} 
 &  & GRPCOVLY & 2 & Other Private \\ \cline{3-5} 
 &  & MRKSCOVLY & 2 & \multirow{3}{*}{NonGroup} \\
 &  & MRKCOVLY & 2 &  \\
 &  & NMCOVLY & 2 &  \\ \cline{3-5} 
 &  & PUBPART & 2 & \multirow{7}{*}{Other Public} \\
 &  & TRCCOVLY & 2 &  \\
 &  & CHAMPVALY & 2 &  \\
 &  & INHCOVLY & 2 &  \\
 &  & PUBCOVLY & 2 &  \\
 &  & HICHAMP & 2 &  \\
 &  & HIMCARE & 2 &  \\ \cline{3-5} 
 &  & HIMCAID & 2 & Medicaid \\ \cline{2-5} 
 & \multirow{13}{*}{MEPS} & INSURCY1 & 3 & Uninsured \\ \cline{3-5} 
 &  & INSURCY1 & 1 & Other Private \\ \cline{3-5} 
 &  & PRSTXY1 & 1 & \multirow{3}{*}{NonGroup} \\
 &  & PRX*1 & 1 &  \\
 &  & PNG*1 & 1 &  \\ \cline{3-5} 
 &  & INSURCY1 & 2 & \multirow{6}{*}{Other Public} \\
 &  & INSURCY1 & 4 &  \\
 &  & INSURCY1 & 5 &  \\
 &  & INSURCY1 & 6 &  \\
 &  & INSURCY1 & 7 &  \\
 &  & INSURCY1 & 8 &  \\ \cline{3-5} 
 &  & MCDEVY1 & 1 & \multirow{2}{*}{Medicaid} \\
 &  & MCD*Y1 & 1 &  \\ \cline{1-5}
\multirow{9}{*}{HealthStatus} & \multirow{9}{*}{MEPS} & \multirow{9}{*}{RTHLTH3} & -9 & \multirow{4}{*}{N/A} \\
 &  &  & -8 &  \\
 &  &  & -7 &  \\
 &  &  & -1 &  \\ \cline{4-5} 
 &  &  & 1 & \multirow{3}{*}{Good} \\
 &  &  & 2 &  \\
 &  &  & 3 &  \\ \cline{4-5} 
 &  &  & 4 & \multirow{2}{*}{Bad} \\
 &  &  & 5 &  \\
\bottomrule
\end{longtable}

\section{Appendix B: Participation and spending for case study policy over time}

In the following table, we show the number of participants in the proposed case study policy at the end of years 1, 5, 10, and 15 by gender, race, and age group, and provide mean family income. These projections demonstrate stability in the demographics of the model population affected by the policy over the projected time frame, absent external forces such as behavioral shifts.

\renewcommand{\arraystretch}{1.2} 

\begin{longtable}{lccccc}
\caption{Plan participation and spending at the end of years 1, 5, 10, 15}
\label{tab:10planparticipants}\\
\toprule
\textbf{} & \textbf{Year 1} & \textbf{Year 5} & \textbf{Year 10} & \textbf{Year 15} \\
\hline
\endfirsthead

{Number of Individuals (millions)} & 43.5 & 43.6 & 45.7 & 46.8 \\
\hline
{\% Female} & 50\% & 51\% & 50\% & 50\% \\
\hline
{\% Male} & 50\% & 49\% & 50\% & 50\% \\
\hline
{\% White} & 53\% & 52\% & 52\% & 50\% \\
\hline
{\% Black} & 16\% & 16\% & 15\% & 16\% \\
\hline
{\% Hispanic} & 24\% & 24\% & 25\% & 26\% \\
\hline
{\% < 19} & 29\% & 30\% & 30\% & 29\% \\
\hline
{\% 19 to 34} & 26\% & 25\% & 25\% & 24\% \\
\hline
{\% 35 to 49} & 22\% & 23\% & 22\% & 23\% \\
\hline
{\% 50 to 64} & 23\% & 22\% & 21\% & 23\% \\
\hline
{Mean Family Income (\$ thousands)} & 86.8 & 90.5 & 89.7 & 88.0 \\
\bottomrule
\end{longtable}

\bibliographystyle{unsrtnat}  
\bibliography{LHIEM_bib}

\end{document}